\documentstyle[12pt]{article}
\def\bc{\begin{center}}
\def\ec{\end{center}}
\def\be{\begin{equation}}
\def\ee{\end{equation}}
\def\bea{\begin{eqnarray}}
\def\eea{\end{eqnarray}}

\def\simge{\ \lower-
1.2pt\vbox{\hbox{\rlap{$>$}\lower5pt
\vbox{\hbox{$\sim$}}}}\ }
\begin{document}
\pagestyle{empty} 
\vspace{-0.6in}
\begin{flushright}
\end{flushright}
\vskip 2.0in
\centerline{\large {\bf{Exploring the 
Light-Cone}}}
\centerline{\large {\bf {through}}}
\centerline{\large {\bf {Semi-Inclusive Hadronic 
Distributions}}}
\vskip 1.0cm
\centerline{M. Testa$^{1,2}$}
\centerline{\small $^1$  Theory Division, 
CERN, 1211 Geneva 23,
Switzerland$^{\star}$.}
\centerline{\small $^2$ Dipartimento di 
Fisica, Universit\`a di Roma ``La
Sapienza"}
\centerline{\small Sezione INFN di Roma}
\centerline{\small P.le A. Moro 2, 00185 
Roma, Italy$^{\star}$$^{\star}$.}
\vskip 1.0in
\abstract{Light-cone dominance is established for a particular set of
semi-inclusive observables in $e^+ e^-$ hadronic annihilation. This allows
to deduce, with a certain degree of rigor, the angular distribution
of hadronic energy from first principles,
without invoking quark-hadron duality.}
\vskip 1.0in
\begin{flushleft} 
\end{flushleft}
\vfill
\noindent \underline{\hspace{2in}}\\
$^{\star}$ Address until August 31st, 1998.

\noindent $^{\star}$$^{\star}$ Permanent address.
\eject
\pagestyle{empty}\clearpage
\setcounter{page}{1}
\pagestyle{plain}
\newpage 
\pagestyle{plain} \setcounter{page}{1}

\section {Introduction}

Perturbative QCD provides a successful approximation scheme for the
description of hadronic processes involving large momentum transfers.
While, originally, quantities reliably computable in perturbation theory
were restricted to those related to various forms of operator
expansion (light-cone or short-distance), subsequently the observables
amenable to perturbative evaluation have been enlarged to encompass
the so-called infrared- and collinear-safe ones\cite{rev}.
Observables related to
short-distance or light-cone singularities, however, still maintain a
privileged status. In fact, while from one side the use of
perturbation theory for their evaluation is justified from first principles, 
on the other side there is also, at least in principle,
the possibility to control pre-leading terms through
the use of the operator product expansion.

$e^+ e^- \rightarrow$hadrons was among the first reactions treated
through operator singularity techniques \cite{had}. Its
peculiar interest lies in the fact that, contrary to the other
classical light-cone dominated reactions, the deep-inelastic scattering,
it does not require the introduction of non-perturbative hadron parameters.

However, as soon as we ask more detailed questions related to the
structure of the hadronic final state, the only
available theoretical instrument is renormalization improved perturbation
theory. Using this technique it has been possible to formulate plausible
arguments showing the jet-like distribution of the produced hadrons\cite{ster}.

In this paper we show that a particular semi-inclusive
observable in $e^+ e^-$ hadron annihilation, first introduced and
discussed in ref.\cite{energy}, can be interpreted in terms of
light-cone singularities of three local operators: two insertions of the
electromagnetic current and the energy-momentum tensor.
Such connection was, in fact, shadowed by the authors
of ref.\cite{energy}, but not exploited by them.

This observable gives rather detailed informations on energy
angular distribution, so that its status of
a light-cone dominated quantity is particularly interesting.

Similar considerations apply to the complete hierarchy of observables
considered in ref.\cite{energy}, but, for simplicity,
we will not address to them explicitly.

Approaches somewhat related to the one presented here are discussed
in refs.\cite{ore},\cite{jet}: in particular some of the ingredients of
the present paper can be found in ref.\cite{jet}. However
the main point here, as contrasted
to refs.\cite{ore},\cite{jet}, is the use of localized observables which,
only, allows the exploration of light-cone singularities.

The use of local observables, however, brings in two basic difficulties:
\begin{itemize}
\item the need of localizing the interaction region, in order to be
able to define its light-cone;

\item the intrinsic non-additivity of local operators.
\end{itemize}

As for the first point, we will use the space-time approach to
scattering, which has been recently shown\cite{unstable} to be
quite successful in the discussion of the properties of unstable states 
in relativistic quantum field theory and we will show that the
connectivity properties of
matrix elements provide a natural way to deal with the second point.

In sections \ref{one} and \ref{two} we will discuss the structure of
the simplest of such observables in terms of hadronic intermediate states, both in the
massive and in the massless situation; in section \ref{three} we will
show its light-cone dominance. In section \ref{four} we present the
conclusions.

\section {Hadronic Analysis} \label{one}

In order not to obscure the exposition with kinematical details
we will treat a schematized problem in
which electrons and positrons are scalar particles which interact with hadrons
through a contact interaction, with an action:
\begin{equation}
S_I= \int d^4x e^{\dagger}(x) e(x) J(x)\label{uno}
\end{equation}
where $e(x)$ denotes the (scalar) electron field, $J(x)$ the
hadronic current and the coupling constant has been reabsorbed in the
definition of the current (for instance $J(x)= g \ \phi ^2 (x)$).

The starting point is the construction of the initial state. It consists
of an $e^+$ and an $e^-$ with wave functions localized
far apart, at large negative times, and overlapping around
the origin of coordinates, around time $t=0$:
\begin{equation}
\vert {in} \rangle =\int 
{d^3p_1d^3p_2f_{\underline p_1}g_{\underline
p_2}\vert {\underline p_1,\underline p_2;in} 
\rangle }\label{due}
\end{equation}

Considering the space-time wave functions $f(\underline x,t)$
and $g(\underline y,t)$ associated to $f_{\underline p}$ and $g_{\underline p}$,
where, e.g.:
\begin{equation}
f(\underline x,t) \equiv \int {{d^3p} \over {\sqrt{(2\pi)^{3} 2\omega_{\underline
p}}}} f_{\underline p} 
e^{-i\omega_{\underline p}t+i{\underline p} \cdot {\underline x}}\label{tre}
\end{equation}
the above requirements amount to say that the supports in space of
$f(\underline x,t)$ and $g(\underline y,t)$ are disjoint for negative time $t$, while
overlapping around time $t=0$.

We will further assume that the momentum spreads of
$f_{\underline p}$ and $g_{\underline p}$ are very narrow, so that
$f_{\underline p}$ is strongly peaked around some given
momentum $ {\underline p}_{0}$ and
$g_{\underline p}$ around momentum $- {\underline p}_{0}$ 
so that the total four-momentum is essentially equal to:
\begin{equation}
Q^\star \approx (2\omega_{{\underline p}_{0}},\underline 0)\label{quattro}
\end{equation}
and we are considering the limit
$\vert {\underline p}_{0} \vert \rightarrow \infty$.

The state $\vert {in}\rangle$ is a superposition of a freely propagating
$e^+ e^-$-pair state:
\begin{equation}
\vert {out} \rangle =\int 
{d^3p_1d^3p_2f_{\underline p_1}g_{\underline
p_2}\vert {\underline p_1,\underline p_2;out} 
\rangle }\label{cinque}
\end{equation}
and a state: 
\begin{equation}
\left| {h} \right\rangle \equiv \left| {in} \right\rangle-
\left| {out} \right\rangle \label{sei}
\end{equation}
in which the interaction 
actually takes place, giving rise to hadron production.
The norm $\langle h \vert {h} \rangle$ of the state defined in eq.(\ref{sei})
is the probability that
hadron production actually occurs in the scattering and is therefore
proportional to the total hadron cross section $\sigma_{e^+ e^- \rightarrow
hadrons}$.

At the lowest order in the interaction described by eq.(\ref{uno}),
$\vert {h} \rangle$ is given by:
\begin{eqnarray}
& & \vert {h} \rangle = i \int 
d^4x d^3p_1 d^3p_2 {{f_{\underline p_1}e^{-ip_1x}}\over {\sqrt{(2\pi)^{3} 2\omega_{\underline
p_1}}}} {{g_{\underline p_2}e^{-ip_2x}}\over {\sqrt{(2\pi)^{3} 2\omega_{\underline
p_2}}}}J(x)\vert {0} \rangle=\nonumber\\
& &=i \int 
d^4x f(x) g(x) J(x)\vert {0} \rangle \equiv \label{sette}\\
& & \equiv i \int 
d^4x F(x) J(x)\vert {0} \rangle \nonumber
\end{eqnarray}
where
\begin{equation}
F(x) \equiv f(x) g(x)\label{settea}
\end{equation}

In view of the assumptions made on $f(x)$ and $g(x)$, $F(x)$ has a
support well localized around the origin of space-time and its Fourier
transform:
\begin{equation}
\tilde{F}(Q) \equiv \int d^4x F(x) e^{iQx} \label{spectrum}
\end{equation}
is narrowly peaked around the four-vector
$Q^\star$ defined in eq.(\ref{quattro}).

The probability of hadron production can then be written as:
\begin{eqnarray}
& & \langle {h} \vert {h} \rangle =
\int 
d^4x d^4y F^*(y)F(x) \langle 0 \vert J(y) J(x)\vert {0} \rangle =\nonumber\\
& & = \int 
d^4x d^4y F^*(y)F(x) \langle 0 \vert J(0) J(x-y)\vert {0} \rangle =\nonumber\\
& & =  \int 
d^4x d^4y F^*(y)F(x+y) \langle 0 \vert J(0) J(x)\vert {0} \rangle =\label{norm}\\
& & =  \int d^4Q \vert \tilde F(Q) \vert^2
\int {{d^4x} \over {(2 \pi)^4}}
e^{iQx}\langle 0 \vert J(0) J(x)\vert {0} \rangle \approx \nonumber \\
& & \approx \int {{d^4x} \over {(2 \pi)^4}}
e^{iQ^\star x}\langle 0 \vert J(0) J(x)\vert {0} \rangle 
\int d^4 Q \vert \tilde F(Q) \vert^2 \equiv \nonumber \\
& & \equiv \Pi (Q^\star{}^2)
\int d^4 Q \vert \tilde F(Q) \vert^2 \nonumber
\end{eqnarray}

In eq.(\ref{norm}) we used the narrow packet approximation in
order to factorize the initial wave function dependence in the factor
$\int d^4Q \vert \tilde F(Q) \vert^2$.

The observable we want to study is related to the expectation
value of the hadronic
energy-momentum tensor evaluated at a space-time point $z$:
\begin{equation}
\langle \theta^{\mu \nu}(z) \rangle \equiv
{{\langle h \vert \theta^{\mu \nu}(z) \vert h \rangle}
\over{\langle h \vert {h}
\rangle}}\equiv {S^{\mu \nu}(z)\over{\langle h \vert h\rangle}} \label{otto}
\end{equation}
where $S^{\mu \nu}(z)$ is defined as:
\begin{equation}
S^{\mu \nu}(z)\equiv \langle h \vert \theta^{\mu \nu}(z) \vert h \rangle=
\int 
d^4x d^4y F^*(y)F(x) \langle 0 \vert J(y) 
\theta^{\mu \nu}(z) J(x)\vert {0} \rangle \label{nove}
\end{equation}
The space-time point $z$ is where (and when) the experimental
counters are placed. We will take it far from the
interaction region, and on its light-cone ($z^2\approx 0$).

Out of $\langle \theta^{\mu \nu}(z) \rangle$ we will construct an observable
which, for large $Q^\star{}^2$, will be dominated by
the short distance region $x \approx y$ and by the light-cone regions
$(z-x)^2 \approx 0$ and $(z-y)^2 \approx 0$.

We start studying how $S^{\mu \nu}(z)$ can be expressed in terms of
hadron quantities, by inserting a double completeness sum over outgoing
hadronic states:
\begin{eqnarray}
& & \langle 0 \vert J(y) \theta^{\mu \nu}(z) J(x)\vert {0}
\rangle=\label{dieci} \\
& & =\sum_n \sum_m 
\langle 0 \vert J(y) \vert n; out \rangle \langle n; out \vert
\theta^{\mu \nu}(z) \vert m; out \rangle
\langle m; out \vert J(x)\vert {0} \rangle \nonumber
\end{eqnarray}

As discussed in detail in the following, it is a general feature of the
Haag-Ruelle scattering theory\cite{haag} that, for
asymptotically large $z$, the semi-disconnected
parts of the matrix element
$\langle n; out \vert \theta^{\mu \nu}(z) \vert m; out \rangle$ will
dominate in eqs.(\ref{nove}),(\ref{dieci}). Qualitatively this is due
to the fact that, in the very distant future the outgoing hadronic state
will look like a bunch
of sparse, practically free, particles. We also notice that,
due to the convolution with $F^*(y)F(x)$ in eq.(\ref{nove}),
the total momentum transfer between $\vert n; out \rangle$ and
$\vert m; out \rangle$  is of the order of the 
energy-momentum spread in $\tilde F(Q)$, so that
$\theta^{\mu \nu}(z)$ carries essentially zero four-momentum.

Semi-disconnected contributions are of the form:
\begin{eqnarray}
& & \langle p_1', p_2', \dots, p_n'; out \vert \theta^{\mu \nu}(z) \vert
p_1, p_2, \dots, p_n; out \rangle _{(s-d)}=\label{undici}\\
& & =\sum_{h=1}^n
\delta (\underline p_1' -\underline p_1) 
\dots
\langle p_h' \vert \theta^{\mu \nu}(z) \vert
p_h \rangle \dots \delta (\underline p_n' -\underline p_n)\nonumber
\end{eqnarray}

In view of the very small momentum carried by $\theta^{\mu
\nu}(z)$ we can write:
\begin{equation}
\langle p_h';\vert \theta^{\mu \nu}(z) \vert
p_h; \rangle \approx
{{p_h{}^{\mu} p_h{}^{\nu}} \over {(2 \pi)^3 \omega_{\underline p_h}}}
e^{i(p_h'-p_h)z} \label{undicia}
\end{equation}
In eq.(\ref{undicia}) the variation of the matrix element of the
hadronic energy-momentum tensor on the scale of the
wave packet momentum spread was neglected, while keeping the complete
momentum dependence of the rapidly varying exponential factor.

In the following we  will denote by $n$ the ensemble of the
particles in the intermediate state with momenta $p_1\dots p_n$ and by
$\tilde n_h$ the ensemble obtained by $n$ after taking out the particle
$h$, so that $n=\tilde n_h+h$.

From eqs.(\ref{dieci}),(\ref{undici})and(\ref{undicia}) we get:
\begin{eqnarray}
& & S_{(s-d)}^{\mu \nu}(z)=\int 
d^4x d^4y F^*(y)F(x) \sum_{p_1\dots p_n}
\vert\langle 0 \vert J(0) \vert n; out \rangle \vert^2 \times \label{lunga} \\
& & \times \sum_{h=1}^n\sum_{p'_h} e^{-ip_{\tilde n_h}(y-x)}
{{p_h{}^{\mu} p_h{}^{\nu}} \over {(2 \pi)^3 \omega_{\underline p_h}}}
e^{ip'_h(z-y)} e^{-ip_h(z-x)} \nonumber
\end{eqnarray}
After the insertion of a factor $1=\int d^4Q \ \delta^{(4)}(Q-p_n)$,
eq.(\ref{lunga}) becomes:
\begin{eqnarray}
& & S_{(s-d)}^{\mu \nu}(z)=\int d^4Q
\int 
d^4x d^4y F^*(y)F(x) \sum_{p_1\dots p_n}
\vert\langle 0 \vert J(0) \vert n; out \rangle \vert^2 \times\label{buona} \\
& & \times \sum_{h=1}^n\sum_{p'_h}
e^{-ip_{\tilde n_h}(y-x)}
{{p_h{}^{\mu} p_h{}^{\nu}} \over {(2 \pi)^3 \omega_{\underline p_h}}} e^{ip'_h(z-y)}
e^{-ip_h(z-x)} \delta^{(4)}(Q-p_n)=\nonumber\\
& & =\int d^4Q \tilde{F}(Q)
\sum_{p_1\dots p_n} \vert\langle 0 \vert J(0) \vert n; out \rangle \vert^2
\sum_{h=1}^n\sum_{p'_h} \tilde{F}^*(Q+p'_h-p_h) \times\nonumber\\
& & \times
{{p_h{}^{\mu} p_h{}^{\nu}} \over {(2 \pi)^3 \omega_{\underline p_h}}}
e^{i(p'_h-p_h)z}
\delta^{(4)}(Q-p_{\tilde n_h}-p_h) \nonumber
\end{eqnarray}

Since we are interested in observations taking place very far from the
interaction region, $z$ is large and the integrations over
${\underline p_h}$ and $\underline p'_h$ in eq.(\ref{buona}) can be
performed through the stationary-phase method, as discussed in ref.\cite{haag}.

Both integrals on ${\underline p_h}$ and $\underline p'_h$ have the same
stationary point, ${\underline p}^{\star}{}_h$, determined by:
\begin{equation}
\left.{\partial {\omega_{\underline p_h}} \over {\partial p_h{}^{j}}}\right|_{\underline
p^\star{}_h}=
{{(p^\star}_h)^j \over \omega_{\underline p^\star{}_h} } \equiv (v^\star{}_h)^j =
{{z^j} \over {z^0}} \label{phase}
\end{equation}
Eq.(\ref{phase}) simply says that only states with at least one
hadron with the correct velocity $\underline v^\star{}_h$ to go from any
finite region of space-time up to $z$, will contribute to the sum
in eq.(\ref{buona}).

An important point should be remarked: while $\underline v^\star{}_h$
is independent on the mass $m_h$ of the hadron $h$, the corresponding
energy, $\omega_{\underline p^\star{}_h}=
{m_h \over \sqrt{1-(\underline v^\star{}_h)^2}}$, and
momentum, ${\underline p^\star}_h=m_h
{\underline v^\star{}_h \over \sqrt{1-(\underline v^\star{}_h)^2}}$,
are indeed strongly $h$-dependent.

The exponential factors in eq.(\ref{buona}) must now be expanded around 
${\underline p}^{\star}{}_h$. We have, for example:
\begin{eqnarray}
& & \underline p_h = \underline p^\star{}_h+\underline \eta \label{fluct}\\
& & e^{-ip_hz} \approx 
e^{-i{{z^0} \over {2 \omega_{\underline
p^\star{}_h}}}[\underline \eta \cdot \underline \eta -
(\underline \eta \cdot \underline v^\star{}_h)^2]}\label{exp}
\end{eqnarray}

The quadratic form, $\underline \eta \cdot \underline \eta -
(\underline \eta \cdot \underline v^\star{}_h)^2$, in the exponent of
eq.(\ref{exp}), has the eigenvalues:
\begin{eqnarray}
& & \lambda_1=\lambda_2=1 \nonumber \\
& & \lambda_3=1-(\underline v^\star{}_h)^2\label{eigen}
\end{eqnarray}
so that eq.(\ref{buona}) becomes:
\begin{eqnarray}
& & S^{\mu \nu}(z) \approx \nonumber\\
& & \approx 1/(z^0)^3 \int d^4Q \vert \tilde{F}(Q) \vert^2
\sum_{h=1}^n \sum_{\tilde n_h} 
\vert\langle 0 \vert J(0) \vert \tilde n_h+p^\star{}_h; out \rangle \vert^2 \times\nonumber\\
& & \times {{(\omega_{\underline p^\star{}_h})^2} \over {1-(\underline
v^\star{}_h)^2}}
(p^\star{}_h)^{\mu} (p^\star{}_h)^{\nu}
\delta^{(4)}(Q-p_{\tilde n_h}-p^\star{}_h) \approx \label{hadron}\\
& & \approx {{\int d^4Q \vert \tilde{F}(Q) \vert^2} \over {(z^0)^3}} 
\sum_{h=1}^n \sum_{\tilde n_h} 
\vert\langle 0 \vert J(0) \vert \tilde n_h+p^\star{}_h; out \rangle \vert^2 \times\nonumber\\
& & \times {{(\omega_{\underline p^\star{}_h})^2} \over {1-(\underline
v^\star{}_h)^2}}
(p^\star)^{\mu} (p^\star)^{\nu}
\delta^{(4)}(Q^\star-p_{\tilde n_h}-p^\star{}_h)\nonumber 
\end{eqnarray}
where we have used the narrowness
of the initial wave packets and dropped the subscript $(s-d)$
which reminded
us we are considering semi-disconnected contributions.
In fact it should by now be clear why we can rigorously restrict our considerations
to the semi-disconnected contributions, eq.(\ref{undici}): all other
(more-connected) contributions will be depressed by powers of $1/(z^0)^3$ with
respect to the semi-disconnected ones.

Through eq.(\ref{hadron}) we can now determine physical observables.
We can, for example, compute the energy flux through a given portion
$\Sigma$ of the spherical surface of radius $\vert \underline z \vert$,
as:
\be
\Phi_\Sigma (z^0)\equiv \int_\Sigma
\langle \theta^{0 i}(z) \rangle n^i d^2\Sigma
\ee
where:
\be
d^2\Sigma = \vert \underline z \vert^2 d^2\Omega_h
\ee
is the (spherical) surface element of $\Sigma$ around $\underline z$
and $d^2\Omega_h$ is the solid angle
element around $\underline z$ and therefore, by eq.(\ref{phase}),
around $\underline v^\star{}_h$.
For $z^0\approx \vert \underline z \vert$, i.e. on the light-cone of the
interaction region, we have:
\bea
& & \Phi_\Sigma (z^0) \approx
{{1} \over {\Pi (Q^\star{}^2) z^0}} 
\sum_{h=1}^n \sum_{\tilde n_h} \int_\Sigma  d^2\Omega_h
\vert\langle 0 \vert J(0) \vert \tilde n_h+p^\star{}_h; out \rangle \vert^2
\times\label{flux}\\
& & \times {{\omega_{\underline p^\star{}_h} ^4} \over {1-(\underline
v^\star{}_h)^2}}
\delta^{(4)}(Q^\star-p_{\tilde n_h}-p^\star{}_h)\nonumber
\end{eqnarray}
where we are considering the ultra-relativistic limit $(\underline v^\star{}_h)^2 \approx
1$, corresponding to our choice of position and switching on of
the measuring apparatus.
In this situation we do not distinguish between
$\omega_{\underline p^\star{}_h}$ and $\vert \underline p^\star{}_h \vert$.

Eq.(\ref{flux}) can also be written as:
\bea
& & \Phi_\Sigma (z^0)=\label{unsafe}\\
& & ={{1} \over {\Pi (Q^\star{}^2) z^0}} 
\sum_{h=1}^n \sum_{\tilde n_h} \int_\Sigma  d^2\Omega_h
\vert\langle 0 \vert J(0) \vert \tilde n_h+p^\star{}_h; out \rangle \vert^2 
{{\omega_{\underline p^\star{}_h} ^6}
\over {m_h{}^2}}
\delta^{(4)}(Q^\star-p_{\tilde n_h}-p^\star{}_h)\nonumber
\eea
which shows that $\Phi_\Sigma (z^0)$ is not a nice inclusive
quantity. In particular it does not have a smooth massless limit
and is not an I.R. safe quantity.
From a physical point of view it is in fact much more sensible
to integrate the energy flux over some interval of time,
in order to get the total energy going through $\Sigma$ during the
corresponding time interval. More generally, we can integrate
the energy flux over some function $\Lambda(t)$ representing the response of the
physical apparatus:
\be
\Psi_\Sigma (\Lambda)\equiv \int dz^0 \Phi_\Sigma (z^0)
\Lambda(z^0 -T)\label{test}\\
\ee
where $\Lambda(t)$ is well localized around zero and:
\be
T \equiv \vert \underline z\vert \label{time}
\ee
so that the measuring region will still be localized around the light-cone
of the interaction region. A ``perfect'' counter corresponds
to $\Lambda(t)=1$ for $-\epsilon < t < \epsilon$ and $0$ otherwise.
The $z^0$ integration in eq.(\ref{test}) is equivalent to an integration over
the speed $v^\star{}_h$ of the detected hadron:
\begin{eqnarray}
& & v^\star{}_h=\vert \underline z \vert / z^0 \label{velocity}\\
& & d z^0/z^0= v^\star{}_h d(1/v^\star{}_h)\approx - d v^\star{}_h \nonumber
\end{eqnarray}
always in the ultra-relativistic approximation ($v^\star{}_h\approx 1$).

Eq.(\ref{test}) then becomes:
\begin{eqnarray}
& & \Psi_\Sigma (\Lambda)\approx 
{{1} \over {\Pi(Q^\star{}^2)}} \sum_{h=1}^n
\sum_{\tilde n_h} \int_{0}^1 d v_h \int_\Sigma d^2\Omega_h
\vert\langle 0 \vert J(0) \vert \tilde n_h+p_h; out \rangle \vert^2
\times\label{hadron1}\\
& & \times {{\omega_{\underline p_h} ^4} \over {1-(\underline
v_h)^2}}
\delta^{(4)}(Q^\star-p_{\tilde n_h}-p_h)
\Lambda[\vert \underline z\vert(1/v_h-1)]\nonumber
\end{eqnarray}
where we dropped the $\star$ superscript on the momentum and velocity
of the $h$-th hadron, because they are from now on dummy integration variables. 
In eq.(\ref{hadron1}) we used:
\begin{equation}
\Lambda(z^0-T)=\Lambda[\vert \underline z\vert (1/v_h -1)]
\end{equation}

Due to the limited support of $\Lambda(t)$, we see that, for large
$\vert \underline z\vert$, the integral over
$v_h$ in eq.(\ref{hadron1}) is restricted to a very small region just below
$1$, which justifies our ultra-relativistic approximations.

We also observe that:
\be
{dv_h \over {1-(\underline v_h)^2}}\approx
{d\omega_{\underline p_h} \over \omega_{\underline p_h}}
\ee
where $\omega_h \equiv \omega_{\underline p_h}$, so that eq.(\ref{hadron1})
can be rewritten as:
\begin{eqnarray}
& & \Psi_\Sigma (\Lambda)\approx
{{1} \over {\Pi(Q^\star{}^2)}}
\sum_{h=1}^n \sum_{\tilde n_h} \int
\omega_h{}^3 d \omega_h \int_\Sigma d^2\Omega_h 
\vert\langle 0 \vert J(0) \vert \tilde n_h+p_h; out \rangle \vert^2
\times \nonumber \\
& & \times \delta^{(4)}(Q^\star-p_{\tilde n_h}-p_h)
\Lambda[\vert \underline z\vert(1/v_h-1)] \label{hadron3}
\eea
In eq.(\ref{hadron3}) we can now make the identification:
\be
\omega_h{}^2 d \omega_h d^2\Omega_h=
d^3p_h \label{measure}
\ee
where $d^3p_h$ is the integration measure of the detected hadron,
so that:
\bea
& & \Psi_\Sigma (\Lambda)
\approx {{1} \over {\Pi(Q^\star{}^2)}} \sum_{h=1}^n
\sum_{\tilde n_h} \int_\Sigma d^3p_h
\omega_h
\vert\langle 0 \vert J(0) \vert \tilde n_h+p_h; out \rangle \vert^2
\times \label{hadron2} \\
& & \times \delta^{(4)}(Q^\star-p_{\tilde n_h}-p_h)
\Lambda[\vert \underline z\vert(1/v_h-1)]\nonumber \\
& & = {{1} \over {\Pi(Q^\star{}^2)}} \sum_n
\vert\langle 0 \vert J(0) \vert n; out \rangle \vert^2
\delta^{(4)}(Q^\star-p_n) \sum_{h=1}^n \omega_h
\Lambda[\vert \underline z\vert(1/v_h-1)] \theta(\underline p_h
\Rightarrow \Sigma)\nonumber
\end{eqnarray}
where $\theta(\underline p_h \Rightarrow \Sigma)$ is $1$, if
$\underline p_h$ crosses $\Sigma$ and $0$ otherwise, 
and reminds us
that the sum runs over intermediate states
containing at least a hadron with momentum direction contained within
$\Sigma$; besides, in eq.(\ref{hadron2}) the $\underline p_h$
integration region is also limited to hadrons whose velocity is close to
$1$, through the presence of $\Lambda[\vert \underline z\vert(1/v_h-1)]$.
The quantity $\Psi_\Sigma (\Lambda)$
is therefore a measure of the energy transported through $\Sigma$
by the fastest hadrons.

\section {The Massless Limit} \label{two}

We discuss in this section the massless limit of $\Psi_\Sigma (\Lambda)$.
This is important for two reasons:
\begin{itemize}
\item in the massless case the space-time behaviour of correlation functions is
rather different from the one found in section \ref{one} and this
discussion will convince us of the infrared safety of  $\Psi_\Sigma (\Lambda)$;

\item in section \ref{three} $ \Psi_\Sigma (\Lambda)$ will be shown to be light-cone dominated, so that
its leading contribution can be reliably computed in massless perturbation
theory. We must therefore find out to which perturbative quantity, expressed
through massless quarks and gluons, it corresponds.
\end{itemize}

The main difference with the treatment of the massive hadron case, discussed in
section \ref{one},
is that in the massless case the stationary phase condition eq.(\ref{phase})
{\it requires}, for consistency, $z^2=0$ {\it and} does not fix
$(p^\star)^\mu$ completely, but only its direction. In fact the
solution of eq.(\ref{phase}), in the massless case, is:
\be
(p^\star)^\mu= \lambda z^\mu \label{valley}
\ee
with an arbitrary $\lambda \geq 0$. The existence of a continuous line of
stationary phase points results in a zero eigenvalue of the quadratic
part of the exponential
factor, eq.(\ref{exp}), as in fact confirmed by eq.(\ref{eigen}). The
correct strategy to adopt in this case is to integrate exactly along the
"valley", eq.(\ref{valley}), and apply the stationary phase approximation
in the transverse directions. In order to carry on this procedure we choose to
parametrize $(p^\star)^\mu$ as follows:
\be
(p^\star)^\mu=\omega {z^\mu \over \vert \underline z \vert}
\ee
This means that the integrations over
$\underline p_h$ and $\underline p'_h$ will be replaced by two one-dimensional
integrations over the corresponding energies, $\omega_h$ and
${\omega_h}'$. We have, therefore:
\begin{eqnarray}
& & S_{m.l.}{}^{\mu \nu}(z)\approx 
{{1} \over {(z^0)^2}} \int d^4Q \tilde{F}(Q) \sum_{h=1}^n
\sum_{\tilde n_h} \int d \omega_h d \omega_h'
\tilde{F}^*[Q+{p^\star}'{}_h-p^\star{}_h]
 \times\label{massless} \\
& & \times \left|\langle 0 \vert J(0) \vert \tilde n_h+p^\star{}_h;
out \rangle_{m.l.} \right|^2
\omega_h {{ (p^\star{}_h)^{\mu} (p^\star{}_h)^{\nu}} \over {2 \pi}}
e^{i({p^\star{}_h}'-p^\star{}_h)z}
\delta^{(4)}(Q-p_{\tilde n_h}-p^\star{}_h) \nonumber
\end{eqnarray}
where the subscript $m.l.$ reminds us that we are in the massless limit.
It should be noticed, that, due to the zero eigenvalue in eq.(\ref{eigen})
($(\underline v^\star{}_h)^2=1$),
in eq.(\ref{massless}) the overall $(1/z^0)^2$
replaces the $(1/z^0)^3$ behaviour of the massive case of eq.(\ref{hadron}).

Another important difference with respect to
eq.(\ref{hadron}) is that, in eq.(\ref{massless}), the integrals over
$\omega_h$ and $\omega_h'$, along the valley direction,
are regulated by the presence of $\tilde{F}^*(Q+{p^\star{}_h}'-p^\star{}_h)$, so that
the dependence from the initial wave packets cannot be factorized.
Therefore
$S_{m.l.}{}^{\mu \nu}(z)$ will depend on the details of the preparation of
the initial $e^+ e^-$ state. This fact has a physical interpretation:
a sharp space-time observation at $z$, would reveal a light front structure
reflecting the detailed characteristics of the initial beam wave packet.

As discussed in section \ref{one}, a physically more realistic observable
is $\Psi_\Sigma (\Lambda)$,
defined in eq.(\ref{test}), which, in the present case, reads:
\begin{eqnarray}
& & \left. \Pi(Q^\star{}^2) \right|_{m.l.} \int d^4 Q \vert \tilde{F}(Q) \vert ^2
\left. \Psi_\Sigma (\Lambda) \right|_{m.l.}=\label {buonaf} \\
& & = \int d^4Q \tilde{F}(Q) \int_\Sigma d^2\Sigma \sum_{h=1}^n
\sum_{\tilde n_h}
\int d \omega_h d \omega_h' 
\tilde{F}^*(Q+{p^\star{}_h}'-p^\star{}_h) \times\nonumber\\
& & \times \left| \langle 0 \vert J(0) \vert \tilde n_h +p^\star{}_h;
out \rangle_{m.l.} \right|^2
{{\omega_h} \over {(2 \pi)^3 }}
\delta^{(4)}(Q-p_{\tilde n_h}-p^\star{}_h) \times \nonumber \\
& & \times \tilde{\Lambda}
(\omega_h'-\omega_h)
\exp +i[(\omega_h'-\omega_h)T
-(\underline p^\star{}_h{}'-\underline p^\star{}_h)\cdot \underline z]
= \nonumber \\
& & ={{\vert \underline z \vert^2} \over {T^2}} \int d^4Q \tilde{F}(Q)
\sum_{h=1}^n \sum_{\tilde n_h} \int_\Sigma d^2\Omega_h \int
d\omega_h d\omega_h'
\tilde{F}^*[Q+{p_h}'-p_h] \times\nonumber\\
& & \times \left| \langle 0 \vert J(0) \vert \tilde n_h+p_h;
out \rangle_{m.l.} \right|^2
{{(\omega_h)^3 } \over {2 \pi}}
\delta^{(4)}(Q-p_{\tilde n_h}-p_h)
\tilde{\Lambda}({\omega_h}' - \omega_h) \nonumber
\end{eqnarray}
where we dropped again the $\star$ superscript in the last step and
we put the exponential to $1$ because of eq.(\ref{time}).

If the support of the smearing function $\Lambda(t)$
is larger than the overlap time of the initial wave packets, the
corresponding support of its Fourier transform, $\tilde{\Lambda}(\omega)$,
will be smaller than that of $\tilde{F}$,
thus allowing to neglect ${p_h}'-p_h$ in
$\tilde{F}^*[Q+{p_h}'-p_h]$. Therefore we have:
\begin{eqnarray}
& & \left. \Psi_\Sigma (\Lambda) \right|_{m.l.}= \label{buonafff}\\
& & = {{1} \over {\left. \Pi(Q^\star{}^2) \right|_{m.l.}}} 
\sum_{h=1}^n
\sum_{\tilde n_h} \int_\Sigma d^2\Omega_h \int d\omega_h
d{\omega_h}'
\left| \langle 0 \vert J(0) \vert \tilde n_h+p_h;
out \rangle_{m.l.} \right|^2 \times\nonumber\\
& & \times 
{{(\omega_h)^3 } \over {2 \pi}}
\delta^{(4)}(Q^\star-p_{\tilde n_h}-p_h)
\tilde{\Lambda}({\omega_h}'-\omega_h) =\nonumber\\
& & = {{\Lambda(0)} \over {\left. \Pi(Q^\star{}^2) \right|_{m.l.}}} 
\sum_{h=1}^n
\sum_{\tilde n_h} \int_\Sigma d^2\Omega_h \int d\omega_h
\left| \langle 0 \vert J(0) \vert \tilde n_h+p_h;
out \rangle_{m.l.} \right|^2 \times\nonumber\\
& & \times 
(\omega_h)^3
\delta^{(4)}(Q^\star-p_{\tilde n_h}-p_h) \nonumber
\eea
Recalling eq.(\ref{measure}), we get:
\bea
& & \left. \Psi_\Sigma (\Lambda) \right|_{m.l.}= \label{buonaff}\\
& & = {{\Lambda(0)} \over {\left. \Pi(Q^\star{}^2) \right|_{m.l.}}}
\sum_n \left|\langle 0 \vert J(0) \vert n;
out \rangle_{m.l.} \right|^2 
\delta^{(4)}(Q^\star-p_n) \sum_{h=1}^n \omega_h
\theta(\underline p_h \Rightarrow \Sigma) \equiv \nonumber\\
& & \equiv \Lambda(0) E(\Sigma) \nonumber
\end{eqnarray}
where:
\begin{equation}
E(\Sigma)= 1/\left. \Pi(Q^\star{}^2) \right|_{m.l.}
\sum_n \left|\langle 0 \vert J(0) \vert n;
out \rangle_{m.l.} \right|^2 
\delta^{(4)}(Q^\star-p_n) \sum_{h=1}^n \omega_h
\theta(\underline p_h \Rightarrow \Sigma) 
\end{equation}
is precisely the first of a hierarchy of observables studied in
ref.\cite{energy}.

The factorization of $\Lambda(0)$ in eq.(\ref{buonaff}) shows that,
in the massless case,
the hadron shock wave produced from $e^+ e^-$ annihilation reaches the
experimental apparatus quite sharply, at the time $T=\vert \underline z \vert$.
In the case of a ``perfect'' counter $\Lambda(0)=1$ and
$\left. \Psi_\Sigma (\Lambda) \right|_{m.l.}=E(\Sigma)$.

The result of a perturbative computation will have the form given in
eq.(\ref{buonaff}), with intermediate states composed of massless
quarks and gluons. Light-cone dominance, discussed in the next section,
implies that, for large $Q^\star{}^2$, perturbation theory should 
reliably reproduce $\Psi_\Sigma(\Lambda)$. Thus, taken together,
eqs.(\ref{hadron2}) and (\ref{buonaff}) imply that,
as $Q^\star{}^2 \rightarrow \infty$, the hadron
velocity distribution will become more and more peaked around $1$, so that
$\Psi_\Sigma (\Lambda)$ becomes proportional to $\Lambda(0)$, as its perturbative
counterpart. Besides, the average value of the energy of the fastest hadrons
crossing $\Sigma$, i.e. $\Psi_\Sigma (\Lambda)/\Lambda(0)$, should be well approximated
by $E(\Sigma)$. This means, in particular, that the energy angular distribution
of fast hadrons asymptotically coincides with that of the corresponding
perturbative massless quark-gluon matter.

\section {Light Cone Dominance} \label{three}

In order to establish the light-cone dominance of the observable
$\Psi_\Sigma (\Lambda)$ defined in eq.(\ref{test}), we use translational invariance and a change
of variables to write:
\begin{eqnarray}
& & \sum_i \int d\mu_{(\Sigma,\Lambda)}{}^i (z) \ S^{0 i}(z)=\label{correlator}\\
& & =\sum_i \int d\mu_{(\Sigma,\Lambda)}{}^i (z) \ d^4x d^4y F^*(y)F(x) \langle 0 \vert J(y) 
\theta^{0 i}(z) J(x)\vert {0} \rangle=\nonumber \\
& & =\sum_i \int d\mu_{(\Sigma,\Lambda)}{}^i (z) \ d^4x d^4y F^*(y)F(x) \langle 0 \vert J(0) 
\theta^{0 i}(z-y) J(x-y)\vert {0} \rangle= \nonumber\\
& & =\sum_i \int d\mu_{(\Sigma,\Lambda)}{}^i (z+y) \ d^4x d^4y F^*(y)F(x+y) \langle 0 \vert J(0) 
\theta^{0 i}(z) J(x)\vert {0} \rangle \nonumber
\end{eqnarray}
where the integration measure $d\mu_{(\Sigma,\Lambda)}{}^i (z)$ summarizes
the angular and temporal averages described in section \ref{one}. 

Since $z$ is very large we can safely approximate, in eq.(\ref{correlator}),
$d\mu_{(\Sigma,\Lambda)}{}^i (z+y) \approx d\mu_{(\Sigma,\Lambda)}{}^i (z)$, so that
we can factorize the wave function dependence and get:
\be
\Psi_\Sigma (\Lambda)\approx 
{{1} \over {\Pi(Q^\star{}^2)}}
 \sum_i \int d\mu_{(\Sigma,\Lambda)}{}^i (z)  \int {{d^4x} \over {(2 \pi)^4}} e^{-iQ^*x}
\langle 0 \vert J(0) \theta^{0 i}(z) J(x)\vert {0} \rangle \label{correlator1}
\ee
where, once again, the narrowness of the initial wave-packets has been exploited.

Eq.(\ref{correlator1}) explicitly shows the light-cone nature
of $\Psi_\Sigma (\Lambda)$. In fact, since the dependence
from $Q^\star$ is expressed in the form of a Fourier transform,
the large $Q^\star$ limit will be dominated by the most singular regions 
of the integrand, i.e., its simultaneous short-distance and light-cone 
singularities.

\section{Conclusions} \label{four}

We have shown that a particular class of semi-inclusive observables can
be related to light-cone and short-distance singularities of products of local operators.
Although we analyzed in detail the simplest of such observables, similar
considerations apply to the whole set of ``energy-energy correlators''
considered in ref.\cite{energy}.
The operator formulation allows to deduce, with a certain rigor,
the angular distribution of hadronic energy from first principles,
without invoking quark-hadron duality. For example, on the basis of
these arguments, we are able to exclude the possibility that the typical
high $Q^\star$ hadronic event consists of a large number of slowly moving
hadrons. The present approach could also allow a systematic
study of pre-asymptotic contributions coming from pre-leading singularities 
in the sort-distance, light-cone operator product expansion.

\section*{Acknowledgements}

I wish to thank the CERN Theory Division for the kind hospitality.

\noindent I am indebted to G. Altarelli, L. Trentadue,
G. Veneziano for discussions. A special thank goes to P. Nason for discussions
and constructive criticism.

\end{document}